\documentclass[aps,pra,pifont,citesort,twocolumn]{revtex4}
\usepackage[utf8]{inputenc}
\usepackage{graphicx,amsmath,amssymb}
\usepackage{dcolumn}
\usepackage{bm}
\usepackage{times}

\newcommand{\bk}[1]{\left ( #1\right )}
\newcommand{\eq}[1]{\begin{equation} \newline #1 \end{equation}}
\newcommand{\eqn}[1]{\begin{eqnarray} \newline #1 \end{eqnarray}}
\newcommand{\ee}{&=&}

\newcommand{\hs}{\hspace{0.2cm}}

\newcommand{\bra}[1]{\left \langle#1 \right |}
\newcommand{\ket}[1]{\left |#1\right \rangle}
\newcommand{\braket}[2]{\langle#1|#2\rangle}
\newcommand{\EV}[1]{\left < #1 \right >}

\newcommand{\nn}{\nonumber}

\newcommand{\adag}{\hat{a}^{\dag}}
\newcommand{\anih}{\hat{a}}
\newcommand{\x}{\hat{x}}
\newcommand{\p}{\hat{p}}
\newcommand{\com}[2]{\left [ #1,#2 \right]}

\begin{document}

\title{Experimental demonstration of Gaussian protocols for one-sided device-independent quantum key distribution}
\author{Nathan Walk$^{1,3,4,*}$}
\author{Sara Hosseini$^{2}$}
\author{Jiao Geng$^{2}$}
\author{Oliver Thearle$^{2}$}
\author{Jing Yan Haw$^{2}$}
\author{Seiji Armstrong$^{2}$}
\author{Syed M. Assad$^{2}$}
\author{Jiri Janousek$^{2}$}
\author{Timothy C. Ralph$^{1}$}
\author{Thomas Symul$^{2}$}
\author{Howard M. Wiseman$^{3}$}
\author{Ping Koy Lam$^{2}$}

\affiliation{$^{1}$Centre for Quantum Computation and Communication Technology, School of Mathematics and Physics, University of Queensland, St. Lucia, Queensland 4072, Australia}
\affiliation{$^{2}$Centre for Quantum Computation and Communication Technology, Department of Quantum Science, Research School of Physics and Engineering, The Australian National University, Canberra ACT 2601, Australia}
\affiliation{$^{3}$Centre for Quantum Computation and Communication Technology,
Centre for Quantum Dynamics, Griffith University, Brisbane, Queensland 4111, Australia}
\affiliation{$^{4}$Department of Computer Science, University of Oxford, Wolfson Building, Parks Road, Oxford OX1 3QD, United Kingdom}
%
%



\date{Compiled \today}



\begin{abstract}
Nonlocal correlations, a longstanding foundational topic in quantum information, have recently found application as a resource for cryptographic tasks where not all devices are trusted, for example in settings with a highly secure central hub, such as a bank or government department, and less secure satellite stations which are inherently more vulnerable to hardware “hacking" attacks.  The asymmetric phenomena of Einstein-Podolsky-Rosen steering plays a key role in one-sided device-independent quantum key distribution (1sDI-QKD) protocols. In the context of continuous-variable (CV) QKD schemes utilizing Gaussian states and measurements, we identify all protocols that can be 1sDI and their maximum loss tolerance. Surprisingly, this includes a protocol that uses only coherent states. We also establish a direct link between the relevant EPR steering inequality and the secret key rate, further strengthening the relationship between these asymmetric notions of nonlocality and device independence. We experimentally implement both entanglement-based and coherent-state protocols, and measure the correlations necessary for 1sDI key distribution up to an applied loss equivalent to 7.5 km and 3.5 km of optical fiber transmission respectively. We also engage in detailed modelling to understand the limits of our current experiment and the potential for further improvements. The new protocols we uncover apply the cheap and efficient hardware of CVQKD systems in a significantly more secure setting.
\end{abstract}


\maketitle

\section{Introduction}

Quantum mechanics promises many new opportunities for the design of communication networks, providing highly correlated resources such as entangled or even nonlocal states as well as stringent restrictions on the possible knowledge of observables, as exemplified by Heisenberg's uncertainty principle. By considering entropic versions of these uncertainty relations \cite{BiaynickiBirula:1975p8601,Maassen:1988p8608} the intimate connection between entanglement and uncertainty, first uncovered in the seminal work of Einstein, Podolsky and Rosen (EPR) \cite{Einstein:1935p402}, has since begun to be formalised and quantified \cite{Berta:2010p1971}.

Both these features are of value to the would-be cryptographer as they enable protocols in which security is grounded in the laws of quantum physics, with the most celebrated example being quantum key distribution (QKD) \cite{Scarani:2009p378}. The earliest, and most conceptually simple, QKD schemes encode a discrete variable (DV) key in a 2-dimensional Hilbert space \cite{bb84-orig,Ekert:1991p460}. As the optical implementation involves sophisticated techniques such as the generation and detection of single photons, considerable attention has also been devoted to schemes that instead utilise the quadratures of the optical field \cite{Ralph:1999p5546,Hillery:2000p8625,Reid:2000p5545,Grosshans:2002p377,Grosshans:2003p2402} where one has access to deterministic, high efficiency broadband sources and detectors. This approach is more theoretically involved however, as the secret key is now a continuous variable (CV) that is encoded in states living in an infinite dimensional Hilbert space. 

The challenge of realising the full promise of QKD - physically guaranteed security with minimal additional assumptions -     has crystallised into two fronts. In the first place, we desire a lower bound on the extractable secret key length, including the effects of a finite number of transmitted symbols, that allows for an arbitrarily powerful eavesdropper (Eve) \cite{Scarani:2008p388,Tomamichel:2012p7120,Furrer:2012p8365,Furrer:2014eq}. In the second place, we would like to close any gaps that may exist between a theoretical QKD protocol and its practical realisation. Essentially, this is the problem of whether or not the honest parties (Alice and Bob) have correctly characterised their experimental devices. One might expect that these gaps must simply be closed on a case-by-case basis. Indeed, as various loopholes due to mischaracterised devices have been pointed out, they have usually been followed by straightforward methods for their closure. Remarkably, however, it is in-principle possible to rigorously surmount even this challenge by harnessing non-local quantum correlations, and it is this second problem we tackle for the entire Gaussian family of CVQKD protocols. We identify all protocols which can be proven secure in a one-sided device-independent (1sDI) setting, i.e. independent of the devices of either Alice or Bob (but not both), and provide a proof-of-principle experimental demonstration some of the most practical of such protocols

Fully device-independent (DI) protocols allow Eve control over all experimental devices and are closely related to the concept of Bell non-locality and the exclusion of local hidden variable (LHV) models \cite{Barrett:2005vd,Acin:2007p384,Masanes:2011p1898,Hanggi:2010p8328,Barrett:2012tu,Vazirani:2014hu}. These schemes are extremely experimentally challenging as they require the implementation of a detection-loophole-free Bell test \cite{Christensen:2013p8763,Giustina:2013p8198,Hensen:2015dw}. As such they are also out of reach for purely Gaussian protocols as it is impossible to violate a Bell inequality utilising only Gaussian resources \cite{BELL:1986dl}. More recently, an intermediate, asymmetric form of non-locality known as EPR-steering has been classified, which allows Alice or Bob to rule out an LHV explanation of the other parties correlations \cite{Wiseman:2007p2187}. A natural question to ask is whether there exist analogous cryptographic results, where only one parties' devices are untrusted. This possibility, first noted in Ref.~\cite{Tomamichel:2011p461} was subsequently developed to prove the security of experimentally difficult, but feasible, proposals for one-sided device-independent (1sDI) DVQKD protocols which were explicitly linked to the corresponding EPR steering inequality \cite{Branciard:2012p8266}. Note that this should not be confused with the distinct concepts of measurement-device-independent QKD, in which both Alice and Bob use trusted sources to generate a key via an untrusted measurement in the middle \cite{Lo:2012p5322,Braunstein:2012p5321,Pirandola:2015in,Tang:2014ua,Rubenok:2013ip} and semi-device independence in which all devices are untrusted but assumptions are made about the Hilbert space dimension\cite{Pawiowski:2011fr}. 

As with Bell tests, closure of the steering detection-loophole has only recently been achieved in state-of-the-art single photon experiments \cite{Bennet:2012ch,Smith:2011cc,Wittmann:2012dg}. This is in stark contrast to the CV case where detection-loophole free tests have been experimentally feasible for over 20 years \cite{Ou:1992p8247} and very strong violations of steering inequalities have been demonstrated \cite{Steinlechner:2013p8065}. Protocols motivated by these hardware advantages have begun to appear. A direct extension Ref.~\cite{Tomamichel:2011p461} to the infinite-dimensional Hilbert spaces relevant for CV-QKD \cite{Berta:2011p8367} has been applied to propose a discretised 1sDI-CVQKD protocol that also accounted for finite-size effects \cite{Furrer:2012p8365,Furrer:2014eq} and a scheme independent of Bob's devices only, has recently been demonstrated \cite{Gehring:2014wc}. 

In this paper we utilise further advances in entropic uncertainty relations \cite{Berta:2013p8691,Frank:km} to theoretically and experimentally investigate the security of the entire family of 16 Gaussian CVQKD protocols against arbitrary attacks in the asymptotic setting. We identify the 6 protocols, including 2 prepare-and-measure (P\&M) schemes, which can be proven 1sDI and compactly calculate their secret key rates. Remarkably, we show that 1sDI-CVQKD is possible with the cheapest and most practical resource in quantum optics - coherent states. We calculate the ultimate limits for all protocols under realistic decoherence channels and show that while reasonably robust to losses, and hence more practical than their discrete variable counterparts over short to medium distances, all the 1sDI-CVQKD protocols are inherently loss-limited. We also make explicit the connection between the asymmetric forms of nonlocality and device-independent cryptography, with the 1sDI-CVQKD key rates displaying an elegant connection to the relevant EPR-steering parameter, a result not known for the DV protocols. Finally, we experimentally implement several protocols including both P\&M and entanglement-based (EB) schemes, finding varying degrees of robustness to losses and experimental imperfections. The best performing protocols allow equivalent losses of up to 7.5 km of optical fibre transmission. Notably, the coherent state protocol has the poorest theoretical loss tolerance, but its experimental performance lies closest to the theoretical limits, indicating it could well be the most practical candidate for short range 1sDI metropolitan networks.

\section{Results}

\subsection{Entropic uncertainty relations and CVQKD}
The most common CVQKD protocols are the Gaussian protocols which encode information in the quadratures of the optical field, described by operators like $\hat{x} = \sqrt{\frac{\hbar}{2}}(\anih + \adag)$ and $\hat{p} = \sqrt{\frac{\hbar}{2}}i(\adag - \anih),$ where $\anih$ and $\adag$ are bosonic annihilation and creation operators. One can prepare squeezed \cite{Ralph:1999p5546,Hillery:2000p8625} or coherent \cite{Grosshans:2003p2402} states, and measure with either homodyne detection (switching between quadratures) or heterodyne  detection \cite{Weedbrook:2004p380} (where both quadratures are measured simultaneously). One could also use EB schemes where two-mode squeezing is used to create Gaussian EPR-correlated states (EPR states) \cite{Reid:2000p5545}. An equivalence between these EB schemes and the P\&M approaches has been established in a device dependent scenario \cite{Grosshans:2003p526}. The communicating parties, Alice and Bob, can also use either a direct reconciliation (DR) scheme where Alice sends corrections to Bob or a reverse reconciliation (RR) \cite{Grosshans:2002p377} where Bob sends corrections to Alice. This makes for a total of 16 protocols. Only the RR protocols allow for losses above 50\%, although one can also achieve this loss-tolerance via post-selection, which discards some of the keys in order to retain a more correlated subset \cite{Silberhorn:2002p149}. 

Previous works have proved the security of Gaussian CVQKD in the asymptotic limit up to the level of collective attacks, via the Gaussian extremality of relevant quantities \cite{GarciaPatron:2006p381,Navascues:2006p805}. Finally the proofs were raised to the level of the most general coherent attacks by use of the de Finetti theorem adapted to infinite dimensions \cite{Renner:2009p1}, which shows that collective attacks are in fact optimal. Consequently, one can asymptotically lower bound the secret key rate by considering only Gaussian collective attacks. For concreteness, we first consider an RR protocol with EPR states and a secret key extracted from Alice and Bob's homodyne measurements denoted by the random variables $X_{A(B)}$ with outcomes $x_{A(B)}$ which follow probability distributions $p(x_{A(B)})$. Neglecting detector and reconciliation efficiencies for simplicity (we shall include these effects in our final calculations) the asymptotic RR secret key rate is lower bounded by \cite{GarciaPatron:2006p381,Navascues:2006p805},
\eq{K^\lhd \geq I(X_A:X_B) - \chi(X_B:E)\label{k}}
where the left-pointing white triangle denotes the direction of information flow during reconciliation from Bob to Alice. A right pointing triangle would signify direct reconciliation from Alice to Bob.
Here 
$I(X_A:X_B) = H(X_A) - H(X_A|X_B)  \label{iab}$ 
denotes the classical mutual information between Alice and Bob, with $H(X) = -\int dx \hs p(x)\log p(x)$ being the continuous Shannon entropy of the measurement strings and
$\chi(X_B:E) = S(E) - \int dx_B \hs p(x_B) S(E|x_B)\label{ie}$
denotes the Holevo bound with $S(E) = -\mathrm{tr}\bk{\rho_E \log \rho_E}$ the von Neumann entropy and $S(E|B) = S(EB) - S(B)$ the conditional von Neumann entropy of $E$ given $B$. In the case that systems are classical, e.g. $B=X_B$, the von Neumann entropies may be replaced by Shannon entropies. 

One can alternatively analyse the security in terms of the conditional entropy of the observable $\x_B$ from the perspective of a quantum eavesdropper $E$,
\eq{S(X_B|E) = H(X_B) + \int dx_B\hs  p(x_B) S(\rho_E^{x_B}) - S(E)\label{cond}}
where $\rho^{x_B}_E$ is the conditional state of $E$ given measurement outcome $x_B$. 

Writing out the key rate in equation (\ref{k}) in full and comparing with equation (\ref{cond}) we have,
\eqn{K^\lhd &\geq & H(X_B) +  \int dx_B \hs  p(x_B) S(\rho_E^{x_B})\nn\\
 && - \hs S(E) - H(X_B|X_A)\nn\\
\ee S(X_B|E) - H(X_B|X_A) }

Bounding the conditional entropy of an observable is the longstanding goal of the study of entropic uncertainty relations \cite{BiaynickiBirula:1975p8601,Maassen:1988p8608}. For our purposes we require a general tripartite relation, encompassing Alice, Bob and Eve, that holds for continuous quadrature observables in an infinite dimensional Hilbert space (see Supplement 1 for details). Very recently, an appropriate relation bounding the entropy of Bob and Eve regarding the conjugate quadratures of Alice has been derived \cite{agnes,Frank:km,Berta:2013p8691}
\eq{S(X_A|E) + S(P_A|B) \geq \log 2\pi\hbar\label{uc2}}

This entropic uncertainty relation now allows us to bound the eavesdroppers information on the relevant observable. Substituting from equation (\ref{uc2}) and recalling that $S(P_A|B)\leq S(P_A|P_B) = H(P_A|P_B)$ we can write,
\eqn{K^\lhd  &\geq &\nn \log 4\pi - H(X_A|X_B) - S(P_B|A)\\
&\geq & \log 4\pi - H(X_A|X_B) - H(P_A|P_B).\label{kh}}
where we have explicitly set our vacuum noise to equal to one (this corresponds to setting $\hbar = 2$).

Thus we have bounded the secret key by an expression that depends only upon the conditional Shannon entropies which are directly accessible to Alice and Bob. Furthermore one can show via a variational calculation that for any probability distribution $p(x)$, the corresponding Shannon entropy is maximised for a Gaussian distribution of the same variance. In other words, Alice and Bob can bound their secret key rate for this protocol by measuring Bob's conditional variances. Substituting the Shannon entropy for a Gaussian distribution $H_G(x_B|x_A) = \log\sqrt{2\pi e V_{X_B|X_A}}$, where $V_{X_B|X_A} = V_{X_B} - \frac{\EV{X_A X_B}^2}{V_{X_A}}$ is Bob's variance conditioned on Alice's measurement, we arrive at the final expression for the RR key rate
\eq{\label{kv}K^\lhd \geq \log \bk{\frac{2}{e\sqrt{V_{X_B|X_A}V_{P_B|P_A}}}}}
The DR expression is obtained by simply permuting the labels of Alice and Bob.
We note that this expression was also calculated in Ref.~\cite{agnes}, but the proof was incomplete as it relied on the assumption of the applicability of the entropic uncertainty relation. Moreover, it was incorrectly concluded that this method would never predict positive key when applied to coherent state or heterodyne protocols. In fact, the extension of equation (6) to the other Gaussian protocols is straightforward and is given in Supplement 1. 

\subsection{One-sided device-independent CVQKD}
An important benefit of utilising entropic uncertainty relations in QKD proofs is that they lend themselves towards one-sided device-independent (1sDI) protocols \cite{Tomamichel:2011p461,Branciard:2012p8266}. These are relaxed versions of fully the DI schemes \cite{Acin:2007p384,Masanes:2011p1898,Hanggi:2010p8328} in which all devices are untrusted and the security is guaranteed via a detection-loophole-free Bell violation. The only assumptions that need to be made for DI schemes are the security of the stations, the causal independence of the measurement trials and a trusted source of randomness for choosing measurement settings. We adopt the same assumptions here, however it should be noted that recently schemes have appeared that do not require causal independence \cite{Barrett:2012tu,Vazirani:2014hu}.

\begin{figure}[t]
\begin{center}
\includegraphics[width=1\columnwidth]{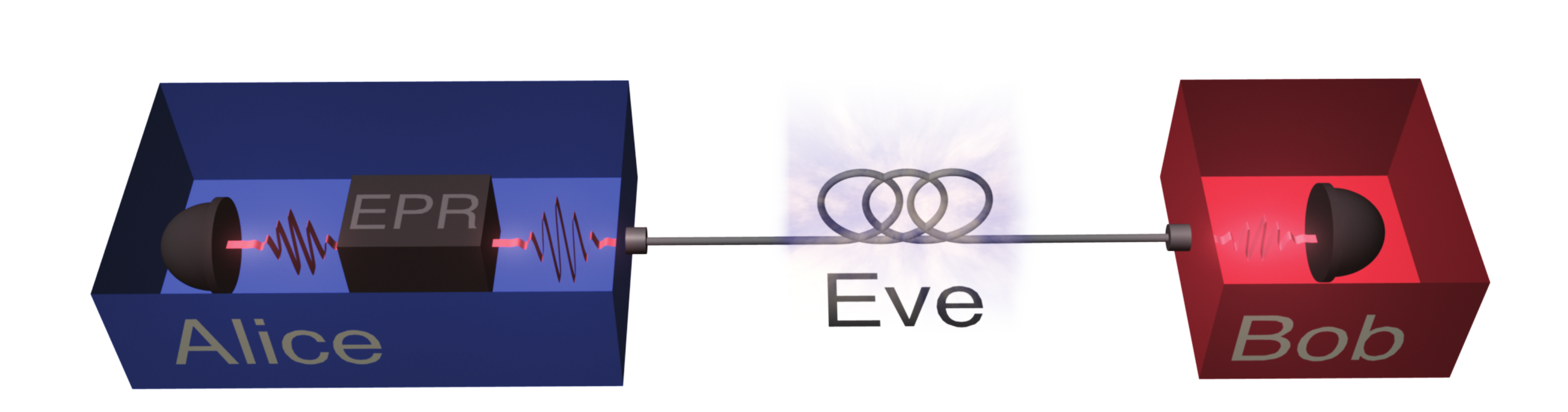}
\caption{Conceptual picture of a 1sDI-CVQKD protocol. From the perspective of Alice (Bob) the local devices are known and allow a secret key to be extracted from a direct (reverse) reconciliation protocol, even though the other party exists only as an unknown red (blue) box.} 
\label{1sdiconcept}
\end{center}
\end{figure}

For 1sDI-QKD protocols only one side, Alice or Bob, is untrusted and regarded as a black box whilst the other is assumed to involve a particular set of quantum operations (see Fig.~\ref{1sdiconcept}). Now, the security is linked to the steering inequalities \cite{Wiseman:2007p2187} associated with the observables on the trusted side. The 1sDI nature of these entropic proofs is manifest in expressions like equation (\ref{kv}) in that it depends only upon measuring a known observable upon one side. For example, in the derivation we only need to know that Bob is measuring either $\x_B$ or $\p_B$ in order to apply the entropic uncertainty relation. Although we write expressions $V_{X_B|X_A}$, as this is what will be measured in experiments, Alice could be making any measurement (not necessarily a quadrature measurement) and the key rate given by equation (\ref{kv}) would still hold. 

\begin{figure}[htbp]
\begin{center}
\includegraphics[width=1\columnwidth]{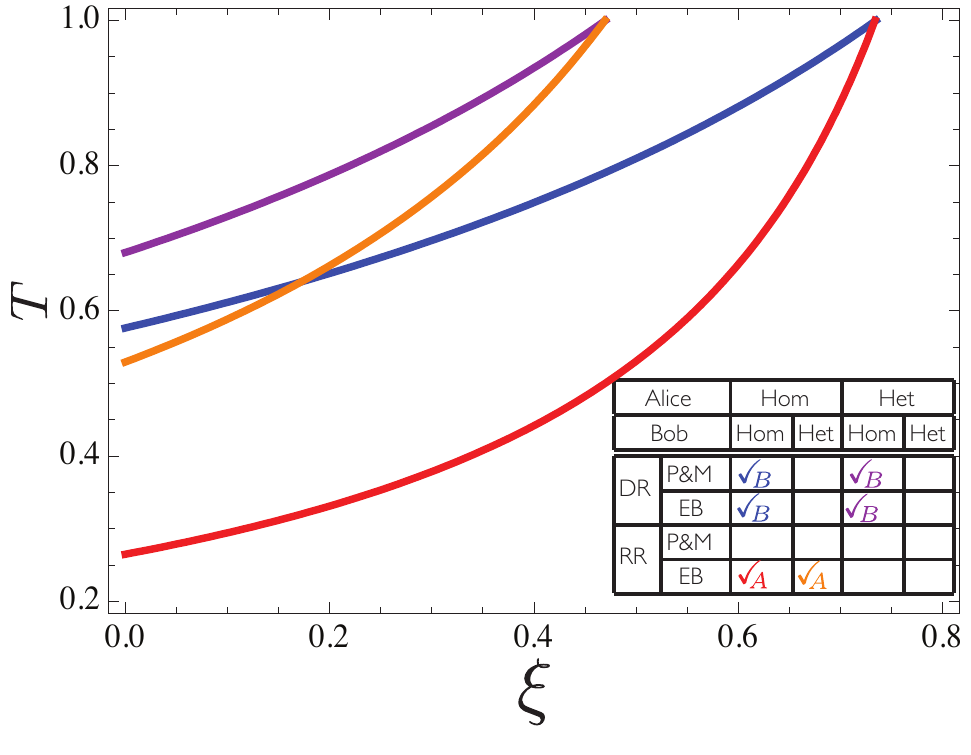}
\caption{Secure regions for 1sDI-CVQKD protocols for a Gaussian channel parameterised by a transmission $T$ and excess noise $\xi$. Direct reconciliation protocols are plotted in blue when Alice homodynes (or equivalently sends squeezed states) and purple when Alice heterodynes (or equivalently sends coherent states). Reverse reconciliation schemes are plotted in red when Bob homodynes and orange when Bob heterodynes. For each protocol secure communication is possible for all channels above the corresponding line. Inset: Summary of 1sDI-CVQKD protocols where subscript A (B) indicates independence of Alice's (Bob's) devices.} 
\label{kth}
\end{center}
\end{figure}

Thus for EPR states and homodyne measurements any positive key predicted via the entropic uncertainty relation is by definition 1sDI, independent of Alice for RR and Bob for DR \cite{Furrer:2012p8365,Furrer:2014eq}. However this device-independence does not necessarily extend to the protocols involving heterodyne detection. This is essentially because the proof to derive non-zero key rates for the heterodyne protocols depends upon characterising the devices used in the heterodyne detection. Therefore, employing a heterodyne detection on the supposedly untrusted side immediately invalidates the device-independence. Alternatively, recall that a steering demonstration requires a measurement choice by the untrusted party \cite{Wiseman:2007p2187} and no such choice takes place if they heterodyne-detect. Nonetheless the remaining protocols, with the heterodyne detection taking place in the trusted station, are still implementable with high-efficiency sources and detection opening the way to several 1sDI-CVQKD protocols with current technology. This means that for EB protocols both DR and RR may be 1sDI provided all parties are homodyning, while Bob may safely heterodyne for an RR protocol and Alice may heterodyne for a DR protocol. Finally, for DR protocols where Alice (who controls the source) is trusted, we may also safely make the equivalence between P\&M and EB schemes. Remarkably, this means that for direct reconciliation it is possible to generate 1sDI key using only coherent states. We summarise which of the 16 possible Gaussian protocols are potentially 1sDI in Table inset in Fig.~\ref{kth}.

Although they allow for 1sDI keys, the entropic proofs result in different, and it turns out generically lower, secret key rates than the standard proofs for Gaussian CVQKD schemes. To map out the ultimate limits of these protocols we first consider an idealised setup with perfect detectors and a highly squeezed (10 dB) two-mode squeezed vacuum source. To evaluate performance, we consider a Gaussian channel, an excellent model for real fibre optic cable, characterised by a transmission $T$ and an excess noise parameter $\xi$ given in units of shot noise. The noise parameter can be thought of as the noise input to the channel such that a pure state with unity variance would have a variance $1+T\xi$ after the channel. Neglecting imperfections such as detector and reconciliation efficiency this is the chief factor that limits the range. In Fig.~\ref{kth} we plot the 4 distinct secure regions (there are two redundancies between P\&M and EB schemes) for the 1sDI protocols. The best performing scheme (in terms of loss tolerance) is the RR EPR scheme where both parties homodyne. In the limit of low excess noise this scheme is secure for up to 73\% loss. For very low noises the next best scheme is the RR protocol where Bob heterodyne detects but for higher noises the DR protocol with both parties homodyning (or alternatively with Alice sending squeezed states) performs better. Finally, although the DR coherent state scheme performs the poorest it is still secure up to around 33\% loss. These results show that 1sDI-CVQKD is reasonably robust to decoherence, but noticeably less loss-tolerant than the standard CVQKD. In this idealised case, the standard protocols tolerate arbitrarily large amounts of loss provided the excess noise is sufficiently small, whereas, our results show that all the Gaussian 1sDI CVQKD protocols are inherently loss limited. Ultimately, this is due to the fact that the uncertainty relations used to bound the secret key rate are only tight when the parties involved, e.g. Bob and Eve in equation (6), can be approximated as sharing a pure, highly-squeezed EPR state \cite{Berta:2013p8691,Furrer:2014eq}. In reality, this is rarely ever the case and the entropic proof method tends to give a pessimistic bound on the eavesdroppers information. We will discuss the prospects for extending the transmission range in the final section.

\subsection{Connection to EPR steering}
\label{subsec:steering}
In the earlier discrete variable work, a clear conceptual link was made between device-independent protocols and Bell non-locality \cite{Acin:2007p384}. Our intuition that the 1sDI-DVQKD protocols should be analogously related to the corresponding asymmetric form of non-locality, EPR steering, was confirmed by Branciard et al. who showed that the condition for their protocol achieving a positive key was equivalent to a steering inequality \cite{Branciard:2012p8266}.

For the Gaussian states and measurements relevant to CVQKD, steering is traditionally demonstrated by a violation of a condition on the conditional variances. In particular, we must violate $ \mathcal{E}_{\blacktriangleright}:=V_{X_B|X_A}V_{P_B|P_A} \geq 1$ for Alice to provably steer Bob as indicated by the right black triangle and similarly with $A$ and $B$ interchanged \cite{Wiseman:2007p2187} and the arrow reversed. This is precisely the same as the EPR paradox criteria derived long ago by Reid \cite{Reid:1989vm}. Comparison with Eq.~(\ref{kv}) shows that we can write the key directly in terms of the steering parameter,
\eqn{K^\lhd \geq \log\bk{\frac{2}{e \sqrt{\mathcal{E}_{\blacktriangleright}}}}}

For the homodyne key rate $K^\lhd>0$ if and only if $\mathcal{E}_{\blacktriangleright}<\bk{\frac{2}{e}}^2\approx 0.55$, with the identical relation between the DR key rate and $ \mathcal{E}_{\blacktriangleleft}$ following straightforwardly. In other words, the condition for a positive one-sided device-independent key is more stringent than EPR steering, similarly to the case for 1sDI-DVQKD \cite{Branciard:2012p8266}. For the protocols where a trusted heterodyne detection takes place, the security of the protocol is instead linked to the steerability of the outcome of the heterodyne measurement which will be more challenging due to the extra loss involved (see Supplement 1). 

Consequently, this connection gives us an operational interpretation for the Reid product of conditional variances \cite{Reid:1989vm} as being directly related to the number of secure 1sDI bits extractable from Gaussian states with Gaussian measurements. This is a particularly practical cryptographic interpretation, in addition to previous work highlighting the links between steering and one-sided device independence in quantum teleportation \cite{Reid:2013vl} and secret sharing \cite{Armstrong:2015he}. Interestingly, the gap between a steering violation and the generation of 1sDI key tells us that Eve's optimal attack (which we know to be Gaussian) followed by a non-Gaussian collective measurement allows her to more effectively steer the Gaussian measurement results of Alice and Bob than any Gaussian measurement.

As a side note, we point out that in the situation where Eve is restricted to individual attacks we would expect a perfect correspondence between steering and key generation since the optimal eavesdropping strategy is known to utilise Gaussian measurements in this scenario. Recalling the secret key formulae when Eve makes Gaussian measurements, $K_G$ \cite{Grosshans:2002p377}, we find this is indeed the case with ,
\eqn{K_G^{\lhd(\rhd)} \geq \log\bk{\frac{1}{\mathcal{E}_{\blacktriangleright(\blacktriangleleft)}}}\label{kg}}
In short, one can also interpret the entropic EPR steering criteria as precisely quantifying the number of secret, 1sDI bits, extractable from a scenario where all parties are restricted to individual measurements.

\subsection{Experimental Results}

\begin{figure*}[ht]

\includegraphics[width=2\columnwidth]{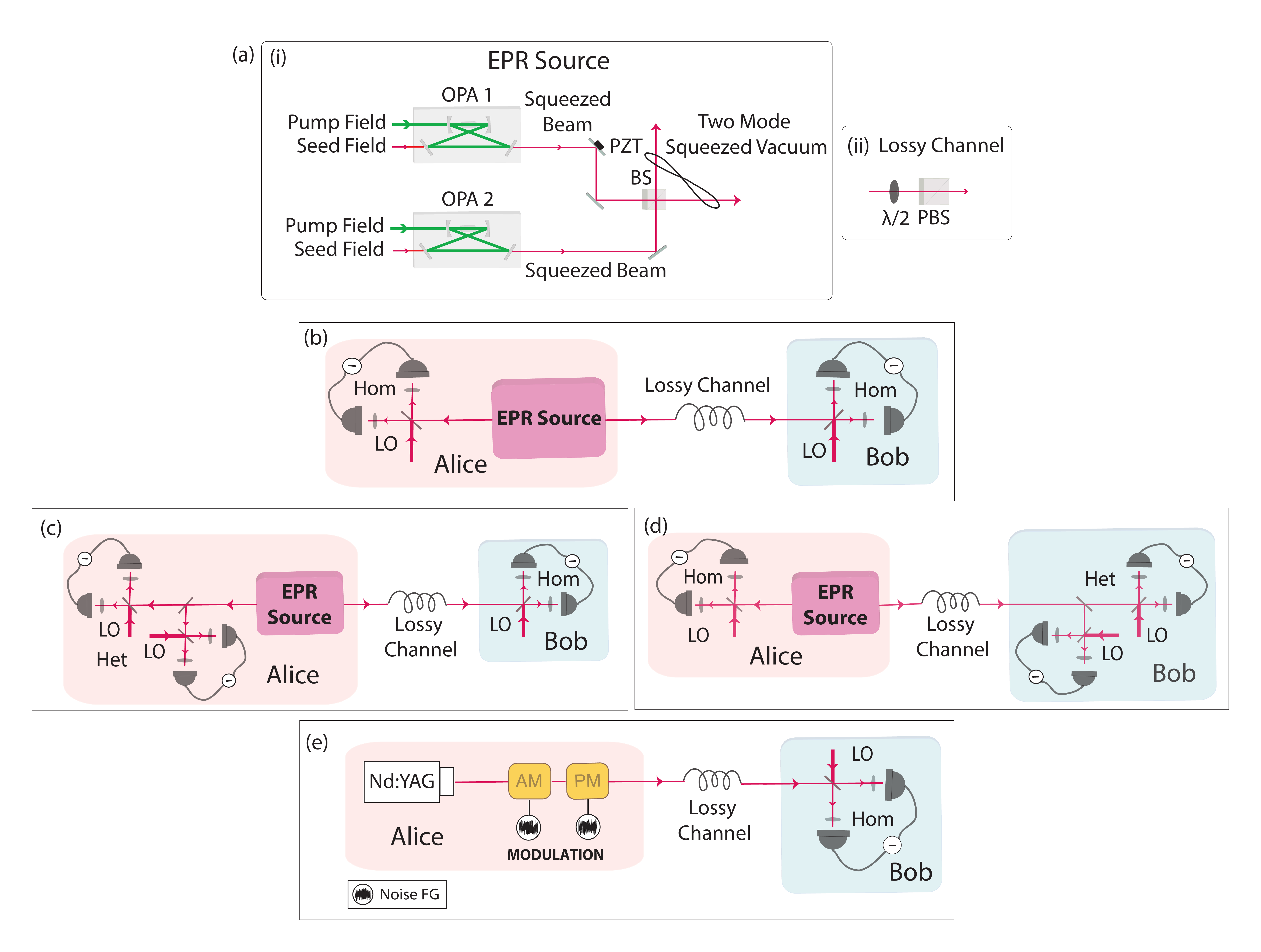}
\caption{Schematic diagram of all the experimental setups. (a)(i) EPR source: OPA1 and OPA2 are two similar optical parametric amplifiers (OPAs) which produced amplitude-squeezed beams. A 1064nm Nd:YAG laser was used to seed both OPAs. They both generated -6.5 dB of squeezing and 10.7 dB of anti-squeezing. PZT is a piezo-electric crystal and BS is a 50:50 beamsplitter. Two amplitude squeezed beams were mixed on a beamsplitter with their relative phase locked in quadrature to produce an EPR state. (a)(ii) optical components used to simulate the lossy channel, which consisted of a half wave plate and a polarizing beamsplitter (PBS). In (b), (c) and (d) one part of the entangled state is sent to Alice locally and the other through a lossy channel to Bob. Here ``Hom" refers to alternating homodyne measurements, and ``Het" to a heterodyne measurement. The measurements configurations for Alice and Bob are (b) Homodyne (Alice) - Homodyne (Bob), (c) Heterodyne (Alice) - Homodyne (Bob) and (d) Homodyne (Alice) - Heterodyne (Bob). (e) P\&M experiment: AM and PM are electro-optic modulators (EOMs) driven by function generators (FG), which in turn provided a Gaussian distributed displacement of the vacuum state in amplitude and phase quadratures. The resulting coherent states were then sent to Bob through a lossy channel where he performed a homodyne measurement. Details of the experimental setups are presented in Supplement 1.}
\label{fig:Sch}
\end{figure*}

As mentioned in the previous sections, 6 of the 16 possible Gaussian protocols are 1sDI. We implement 5 protocols experimentally, 3 of which exhibit sufficient correlations to allow for 1sDI-CVQKD. Two different experimental setups were used, the first for the EB protocols based on EPR correlations and the second for a coherent-state P\&M protocol (see Fig.~\ref{fig:Sch}). To perform the 1st and 2nd 1sDI protocols we used the EPR source while both parties performed a homodyne detection. The 3rd and 4th protocols were implemented using the same EPR source while one party (Alice for the DR protocol and Bob for the RR protocol) heterodyned whilst the other homodyned. Finally, the P\&M scheme was implemented for the DR protocol where Alice, who was trusted and controlled the source, generated coherent states and Bob performed a homodyne measurement. 

\begin{figure}[t]
\begin{center}
\includegraphics[width=1\columnwidth]{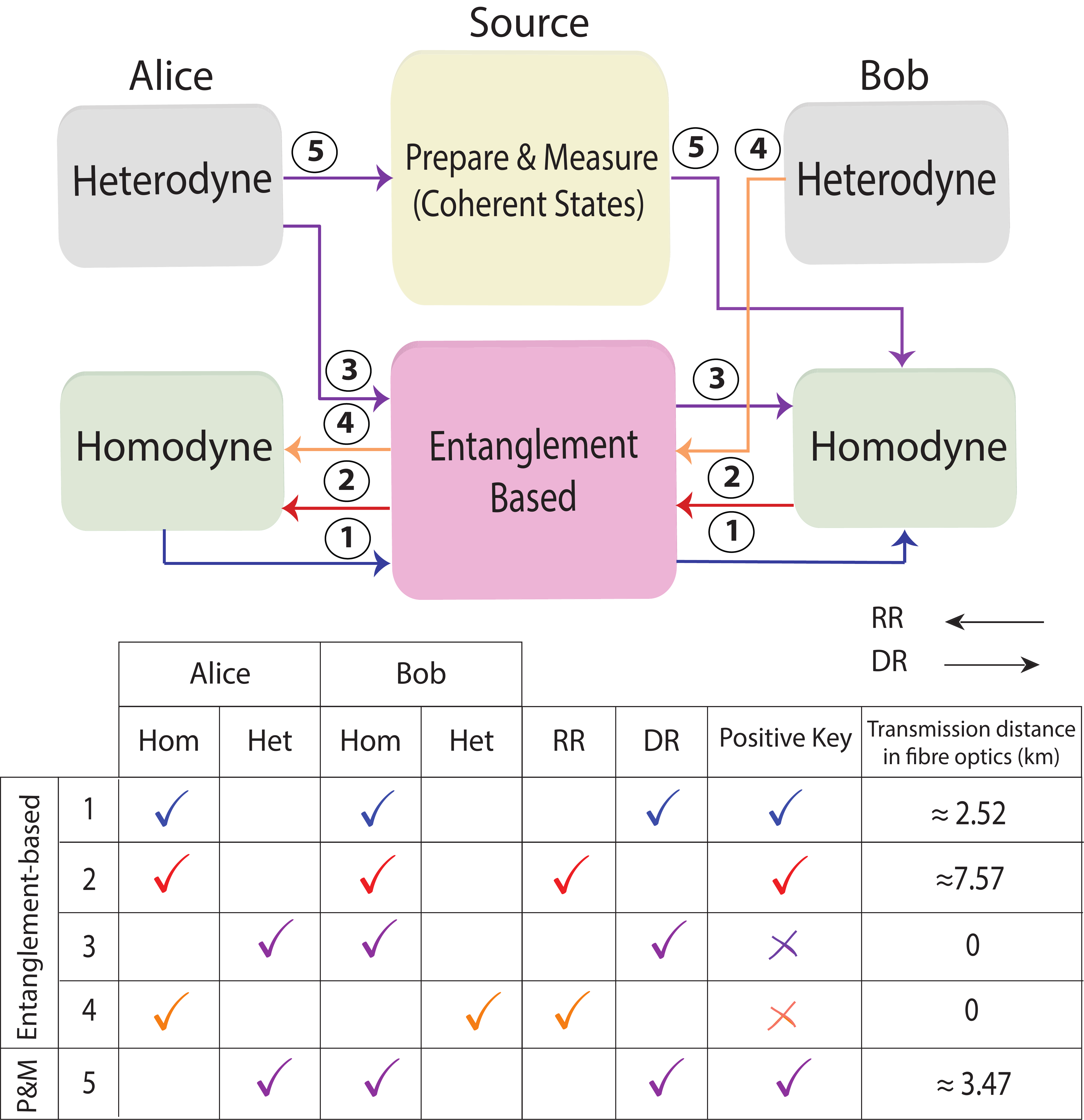}

\caption{Schematic diagram of all the experimentally realised 1sDI protocols. Alice and Bob can choose between homodyne and heterodyne measurements, using a source that generates either EPR or coherent states. Direct (reverse) reconciliation protocols are demonstrated using right (left) pointing arrows. The table summarizes each performed protocol and the experimentally achieved results. The same color scheme as Fig.~\ref{kth} is used here to show the different performed protocols.}
\label{fig:exp}
\end{center}
\end{figure}

In each protocol, Alice and Bob are connected by a lossy channel of transmission $T$. The lossy channel is constructed using a half wave plate and a polarizing beam splitter as detailed in Fig. \ref{fig:Sch} (a)(ii). We express the applied loss as the equivalent transmission distance through a standard telecom optical fibre with a loss of $0.2\mathrm{dB/km}$. Ideally, the secret key rate could be computed directly from the expressions in Supplement 1. However, in practice we must modify these expression, multiplying Alice and Bob's mutual information by a factor $\beta <1$ to account for finite information reconciliation efficiency (see Supplement 1 for explicit calculations). Reconciliation efficiencies for CVQKD have increased substantially in the last few years \cite{Jouguet:2011p5057,Jouguet:2013p8197}, with efficiencies of between 94 and 95.5 percent recently reported \cite{Gehring:2014wc}. Here, we choose $\beta = 0.95$. The inclusion of $\beta<1$ will reduce the final calculated key rate. This makes the condition $\mathcal{E}_{\blacktriangleright(\blacktriangleleft)}< 0.55$ necessary but no longer sufficient for a positive key when $\beta$ is included. A schematic diagram of all the performed experiments and the achieved results are summarized in Fig.~\ref{fig:exp}.

Amongst the successful implementations, protocol 2 (EB scheme RR protocol where both parties homodyned) shows the best loss tolerance and protocol 1 (EB scheme DR protocol where both parties homodyned) shows the worst, with protocol 5 (coherent state P\&M scheme with homodyne detection) being intermediate. This actually demonstrates a different hierarchy of loss tolerance than the theoretical results calculated in the limit of very large squeezing and pure entanglement (Fig.~\ref{kth}). This difference is due to the fact that in our experiment, we only had about -6 dB of squeezing and 10.7 dB of anti-squeezing which, along with other losses and imperfections, degraded the quality of the entangled source and hence limits the range of the EB protocols. This is also the reason why the heterodyne protocols (3 and 4) fail to produce any positive key at all, as overcoming the shot noise penalty requires extremely strong correlations. Our calculations show that a perfect system with no losses of any kind and reconciliation efficiency of 0.95 would still require at least 7dB of perfectly pure squeezing (-7dB squeezing and 7dB anti-squeezing) to get a positive key rate even over a perfect channel with the heterodyne protocols.  

\begin{figure*}[t]
\centering
\includegraphics[width=11cm]{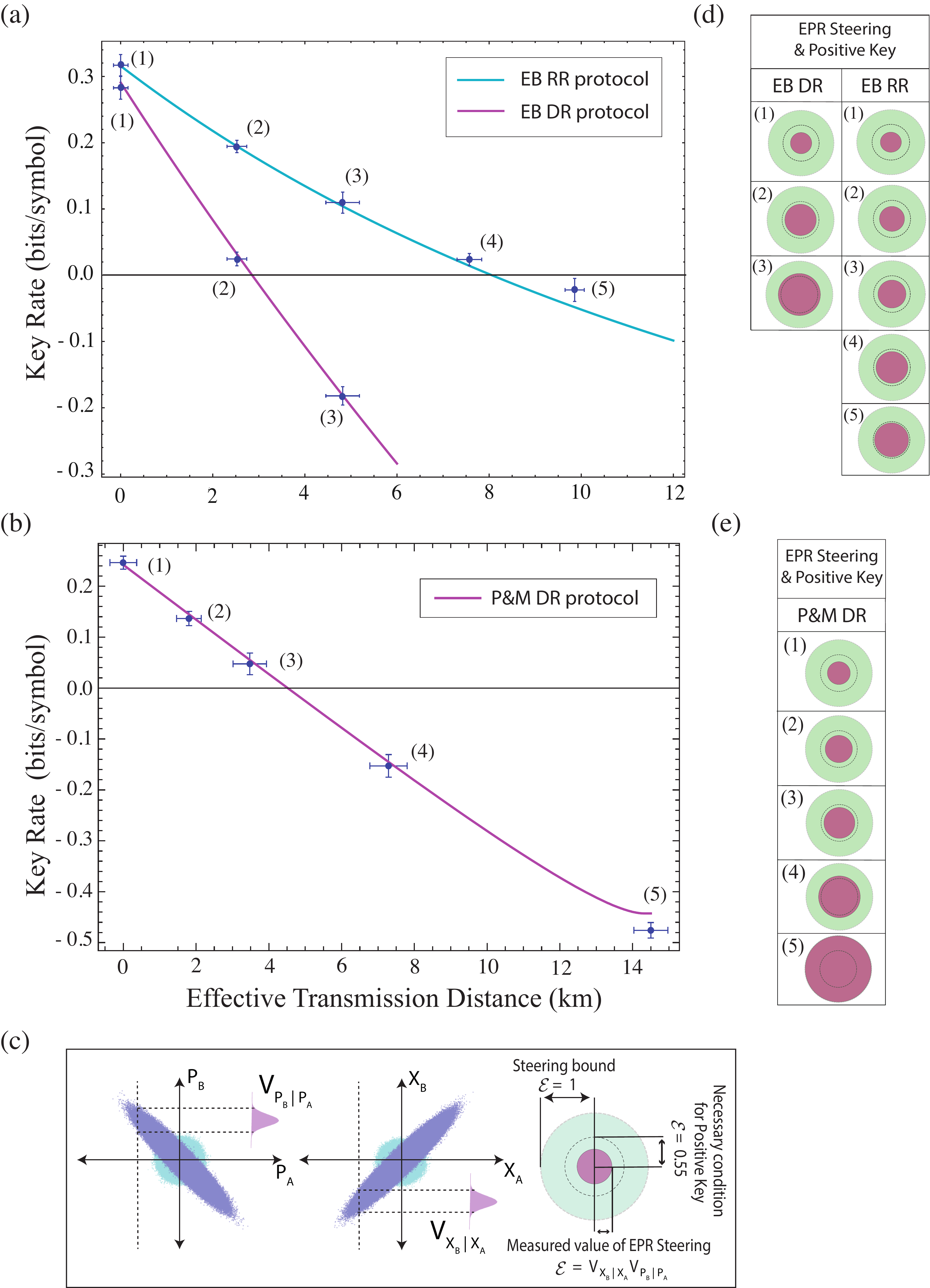}
\caption{Key rates versus distance for (a) DR and RR protocols with EB source and homodyne-homodyne detections and (b) P\&M coherent state DR scheme (protocols 1, 2 and 5 in the Table of Fig.\ref{fig:exp}). Theoretical curves, including imperfect reconciliation efficiency were evaluated from the expressions given in Supplement 1, part 3 using models described in Supplement 1, part 4 and part 5. which took into account the finite reconciliation efficiency $\beta$, source and locking imperfection, and optical losses. The experimental data points are superposed on the plot, with error bars (1 s.d.) estimated from error propagation of uncertainties.  (c) Example experimental data points from the EB DR protocol. Phase-space plots show the correlations between the quadratures, $P_A$ and $P_B$ and $X_A$ and $X_B$, measured by Alice and Bob. Using their statistics, the conditional variances are calculated and used to estimate the EPR-steering parameters. The circle on the right illustrates the comparison of the measured value of the EPR steering (purple), the necessary condition for obtaining a positive key rate (dashed) and the upper bound for EPR steerability (green). Panels (d) \& (e) illustrate circles corresponding to each plotted data point in (a) \& (b), showing the connection between the measured values of EPR steering and the generation of positive key rates.
}
\label{fig:Hom}
\end{figure*}

We plot our measured secret key rates as a function of effective transmission distance in Fig.~\ref{fig:Hom}. Solid lines are calculated from a theoretical model based upon the characterisation of various imperfections in the experiment. Results for the protocols where Alice and Bob performed the homodyne measurements on a distributed EPR state, are given in Fig.~\ref{fig:Hom} (a). Using the RR protocol we measured a positive key rate independent of Alice's devices up to an equivalent transmission distance of 7.57$\pm$ 0.26 km (approximately 29\% applied loss). Using the DR protocol, we measured a secret key independent of Bob's devices up to an equivalent transmission distance of 2.52$\pm $0.21 km (approximately 11\% applied loss). Our theoretical model, which is in good agreement with the experimental data, predicts a maximum transmission range of 8 km and 2.8 km for the RR and DR protocols respectively (see Supplement 1). 

Fig.\ref{fig:Hom} (b) depicts the results of the DR coherent state protocol. We show that secure key remains possible after an equivalent transmission distance of 3.47$\pm $0.46 km (approximately 15\% applied loss). This is in good agreement with our theoretical model, which predicts our current setup would be secure up to a maximum of 4.5 km. With the P\&M protocol, described in Fig.\ref{fig:Sch} (e), we have much more latitude to vary the modulation variance and hence the virtual entanglement in order to optimize the secret key rate for each loss setting (see Supplement 1). As such, we achieve a loss tolerance superior to the EB DR protocol, whilst using only the cheapest and most readily available quantum optical resource states. 

We also display the behaviour of the measured steering parameter with respect to the thresholds required for key generation and violation of the Reid EPR-steering criteria (see Fig.\ref{fig:Hom} (c)). For each data point, we graphically represent the relevant steering parameter with respect to these thresholds in Fig \ref{fig:Hom} (d) \& (e). In accordance with our earlier discussion we show that a positive key is achieved with an EB RR (DR) protocol only when $\mathcal{E}_{\blacktriangleright(\blacktriangleleft)}<\bk{\frac{2}{e}}^2\approx 0.55$. We note that for the P\&M DR protocol, since in the equivalent EB picture Alice performs a heterodyne detection, $\mathcal{E}_{\blacktriangleright}=V_{X_{A_1}|X_B}V_{P_{A_2}|P_B}$. Here, $A_1$ and $A_2$ are the modes upon which Alice measured $\hat{x}$ and $\hat{p}$, respectively. On the other hand, all plotted points demonstrate EPR-steering through a violation of the Reid criteria. Consequently, the negative data points would demonstrate sufficient correlations for 1sDI key generation if we were able to restrict Eve to individual attacks as per equation (\ref{kg}).

In order to better understand the limitations of, and potential improvements to, our experiments, theoretical models of both the EB and P\&M were constructed. Modelling all processes as Gaussian, allows for compact calculations and matches the experimental data closely. As well as using these models to determine the maximum range of our current experiment, we also investigated the performance that could be achieved with an improved implementation. For the EB protocols the dominant source of decoherence are losses in the squeezing cavities. Modelling indicates that making challenging but reasonable improvements to the amount of available squeezing and the precision of locking could extend the asymptotic range of the DR and RR homodyne protocols to around 8 and 16 km respectively, again assuming a reconciliation efficiency of $\beta = 0.95$. See Supplement 1 for detailed explanation of the model for the EB protocols.

As mentioned previously, in contrast to the EB protocols when using coherent states we have a great deal of flexibility in tuning the virtual squeezing via an increase in the modulation strength. In this protocol, the dominant source of noise is the unwanted cross modulation between the quadratures that worsens as the modulation strength increases. This can be thought of as an unknown phase space rotation, and means we cannot use the modulation variance that would otherwise be optimal and depend only upon $\beta$ and the channel loss. If this cross modulation could be eliminated, our model shows the asymptotic range of the coherent state scheme would increase to around 4.5 km. Details of the P\&M model and the modulation optimisation can be found in Supplement 1.


\section{Discussion}

To summarise, we have provided a complete taxonomy of the Gaussian CVQKD protocols from the perspective of one-sided device-independence. We also derived the asymptotic secret key rate for all 6 such protocols, and made an explicit connection to the EPR steering parameters for Gaussian states and measurements. Using these derived rates we have characterised an experimental implementation of 5 of the 6 protocols, achieving secure key under a lossy channel equivalent of up to 7.5 km of optical fibre. Of particular interest was the first demonstration of a 1sDI CVQKD protocol using only coherent states. That such an exotic quantum communication protocol is possible with these relatively mundane quantum states is a surprising result in itself. Furthermore, the ease with which they can be generated makes them an especially attractive candidate for short range metropolitan networks.

Several comments on extensions and directions for future work are in order, beginning with the prospect of extending this security proof to include finite size effects and comparison with the results in Refs.~\cite{Furrer:2012p8365,Gehring:2014wc,Furrer:2014eq}. In particular, in the experiment in ref.~\cite{Gehring:2014wc}, the authors follow a similar program of applying entropic uncertainty relations, in this case to the smooth min-entropies, allowing them to account for all finite size effects while providing proof against completely general attacks whilst implementing one of the protocols described here (protocol 1). With squeezing level of 10 dB, the key rate demonstrated was about 0.1 bit per sample at a distance of 2.7 km, whereby for our setup the range is about 1.6 km, or up to 6 km with comparable squeezing level (see Supplement 1). Nevertheless, this proof is only for DR homodyne protocols and limited to short distances (up to 5 km) even with extremely high levels of squeezing. Very recently, an extension to RR homodyne protocols, for both the asymptotic and finite-size regimes, secure up to 15 km has also appeared \cite{Furrer:2014eq}. For a coherent state homodyne protocol like that discussed here, a finite-size proof has also been developed \cite{Leverrier:2015he}. It seems very promising then, that these techniques could be adapted to prove the finite-size security of all the other protocols presented here. Nonetheless, our asymptotic analysis shows that even in the most ideal situations, 1sDI-CVQKD is presently limited to transmission through urban networks.
%
%
%

An obvious avenue for future work is the investigation of methods to improve long distance performance. One option would be to revisit the restrictions, or lack thereof, made about the eavesdropper including physical assumptions about the quantum memory available to Eve \cite{Wehner:2008p8823,Wehner:2010p8822,Schaffner:2010p9239}, which has already seen application in DI-DVQKD \cite{Pironio:2013p8689}. Another candidate to further extend the range of these protocols would be the noiseless linear amplifier \cite{proc-disc-2009,Xiang:2010p1449} which has already been proposed for application to fully DI-DVQKD \cite{Gisin:2010p7352}. Even more appealing may be the measurement-based versions of these amplification schemes \cite{Walk:2013p403,Fiurasek:2012p7670} that have recently been experimentally demonstrated \cite{Chrzanowski:2014iga} although this could only be applied to RR protocols. In light of these results it appears that several 1sDI-CVQKD protocols are within the reach of current technology and multiple possibilities exist to extend the secure range of such schemes to long distances.

\section*{Funding Information}
ARC Centre of Excellence CE110001027. N.W. acknowledges support from the EPSRC National Quantum Technology Hub in Networked Quantum Information Technologies.

\section*{Acknowledgments}
The authors would like to thank D.A. Evans, M.J. Hall, C. Branciard, E.G. Cavalcanti, H.M. Chrzanowski, F. Furrer and M. Tomamichel for helpful discussions.

\begin{appendix}

\section{Entropic uncertainty relations }
Entropic relations have received a great deal of attention as a convenient and powerful information theoretic tool for investigating uncertainty in quantum systems. Originally, entropic uncertainty relations were derived assuming one starts without any additional information or at most only classical information describing the system in question, i.e. the density matrix \cite{BiaynickiBirula:1975p8601,Maassen:1988p8608}. In either case, since classical information can be shared perfectly amongst arbitrarily many parties, there is little sense in thinking about these relations as applying from the perspective of one observer or another. Conversely, if observers were to share quantum correlations with the measured system, one expects the uncertainty relations to be strongly observer dependent and potentially exhibit reduced levels of uncertainty.

A generalised relation, allowing for this so-called quantum side information, was derived in \cite{Berta:2010p1971} although only for finite dimensional Hilbert spaces and observables with a discrete spectrum. Consider a pair of observables $\{\x_A, \p_A\}$ with a complementarity $c = \max_{p_A,x_A} |\braket{x_A}{p_A}|^2$ where $\{\ket{x_A},\ket{p_A}\}$ are the eigenvectors of the observables. These observables are to be measured on a state $A$ which is potentially entangled with another state, $B$, leading to the the following relation for the uncertainty in the pair of observables given access to $B$ \cite{Berta:2010p1971},
\eq{S(X_A|B)+S(P_A|B) \geq \log \frac{1}{c} + S(A|B).\label{uc1}}
Here $S(A|B) = S(AB) - S(B)$ where $S(X) = -\mathrm{tr}\bk{\rho_X \log \rho_X}$ is the conditional von Neumann entropy of the state $\rho_{AB}$ whereas $S(X_A|B)$ is the conditional von Neumann entropy of the random variable, $X_A$, corresponding to the measurement of the \emph{observable} $\hat{x}_A$ on system $A$ given knowledge of system $B$. This is defined as,
\eq{S(X_A|B) = H(X_A) + \sum_{x_A} p(x_A) S(\rho_B^{x_A}) - S(B)\label{cond1}}
with $H(X_A) = -\sum_{x_A} p(x_A)\log p(x_A)$ the Shannon entropy and $\rho_B^{x_A}$ describing Bob's state conditional on Alice obtaining outcome $x_A$. The presence of the conditional entropy $S(A|B)$ in Eq.~\ref{uc1}, which is negative for entangled states, demonstrates both the observer dependence and effect of entanglement in reducing uncertainty.

Preempting applications to quantum key distribution (QKD), one can also consider that the state $\rho_{AB}$ could have suffered some decoherence which is purified by an environment, or eavesdropper, such that $\rho_{AB} = \mathrm{tr}_E\bk{\ket{ABE}\bra{ABE}}$. Using the purity of the overall state (i.e. $S(AB) = S(E)$) one can recast equation (\ref{uc1}) to find \cite{Berta:2010p1971},
\eq{S(X_A|B) + S(P_A|E) \geq \log \frac{1}{c}.\label{eve}}
However, these results are only valid for measurements with a finite number of discrete outcomes made on states living in a finite-dimensional Hilbert space. For the purposes of continuous variable (CV) QKD we will require an uncertainty relation valid for infinite-dimensional Hilbert spaces and continuous-valued measurements. In particular, we are interested in homodyne measurements of the canonically conjugate quadratures $\hat{x} = \sqrt{\frac{\hbar}{2}}(\anih + \adag)$, $\hat{p} = \sqrt{\frac{\hbar}{2}}i(\adag - \anih) $ satisfying $\com{\x}{\p} = i\hbar$ where $\anih$ and $\adag$ are bosonic annihilation and creation operators. 

Just such a relation has been recently developed, building on an earlier result for discrete and finite measurements on infinite dimensional Hilbert spaces \cite{Berta:2011p8367}. This was first extended to countably infinite measurements which could then be applied to a discretised version of a homodyne detection \cite{Berta:2013p8691}. Deriving results for continuous spectra, by taking infinite precision limits of these coarse-grained POVM's, had previously been extensively studied for the Shannon entropies, and an analogous procedure for the quantum conditional von Neumann entropy was utilised by Ferenczi \cite{agnes} and Berta et al. \cite{Berta:2013p8691}  although the former proof is incomplete. An alternative derivation was also provided by Frank and Lieb \cite{Frank:km}. The final result is the following relation for homodyne detection upon infinite dimensional Hilbert spaces \cite{agnes,Frank:km,Berta:2013p8691},
\eq{S(X_A|B) + S(P_A|E) \geq \log 2\pi\hbar\label{uc2}}

\section{Secret key rates for Gaussian protocols}

We will now derive bounds upon the secret key rate for all members of the Gaussian family of CVQKD protocols. In the P\&M setting, one could consider Alice sending either squeezed \cite{Ralph:1999p5546,Hillery:2000p8625} or coherent states \cite{Grosshans:2003p2402} and Bob measuring with either homodyne or heterodyne  \cite{Weedbrook:2004p380,Weedbrook:2006p8395} detection. Each permutation can in turn be used to have Bob try and guess Alice's encoding, called direct reconciliation \cite{Grosshans:2002p377} (DR), or Alice trying to guess Bob's measurement \cite{Grosshans:2003p2402}, called reverse reconciliation (RR). One can also consider entanglement based (EB) schemes, in which Alice distributes one arm of a two-mode squeezed vacuum to Bob \cite{Reid:2000p5545}. In fact, with appropriate rescaling of Alice's data, the EB and P\&M schemes can be seen as equivalent since having Alice heterodyne detect one arm of a two-mode squeezed vacuum corresponds to preparing a coherent state, while a homodyne detection corresponds to squeezed state preparation \cite{Grosshans:2003p526}. In the following, we conduct our analysis in the EB picture, and the  variances appearing are those that would be directly measured in an EB implementation. We will calculate the key rate encoded in the $\x$ basis on a particular run. Overall, the total key rate will be the average of the quantities derived here and the analogous expression for encoding in the $\p$ basis (the exception is the heterodyne-heterodyne protocol where both bases are always simultaneously utilised). Although we calculate all the key rates for completeness, for the protocols involving heterodyne detection only one reconciliation direction will allow for a 1sDI protocol as per the main text.

\subsection{Homodyne-Homodyne (Squeezed states and Homodyne Detection)}
First we turn to the protocols where both Alice and Bob homodyne detect. Here Alice and Bob randomly choose whether they measure the $\x$ or $\p$ basis and only keep those times where they agree. To obtain effective EB data from a P\&M scheme Alice will rescale her modulation signal by dividing $\alpha_{x(p)}=+(-)\sqrt{1-1/V_{x(p)}^2}$ where $V_{x(p)} = V_{S_{x(p)}}+1$ and $V_{S_{x(p)}}$ is the modulation variance of the signal encoded in the respective quadratures. We work in the asymptotic regime $n \rightarrow \infty$ where the optimal attacks are known to be the collective attacks. Considering the case where Bob makes an $\x$ measurement, we can write the secret key rate for the DR protocol as
\cite{GarciaPatron:2006p381,Navascues:2006p805},
\eq{K^\rhd \geq I(X_A:X_B) - \chi(X_A:E)\label{k1}}
where
\eqn{I(X_A:X_B) \ee\nn H(X_B) - H(X_B|X_A)\\
\ee H(X_A) - H(X_A|X_B) \label{iab1}} 
denotes the classical mutual information between Alice and Bob, with $H(X) = -\int dx \hs p(x)\log p(x)$ being the continuous Shannon entropy of the measurement strings and
\eqn{\chi(X_A:E) = S(E) - \int dx_A \hs p(x_A) S(E|x_A)\label{ie1}}
is the Holevo bound.

Substituting equations (\ref{iab1}) and (\ref{ie1}) into (\ref{k1}) and comparing with the continuous conditional von Neumann entropy 
\eq{S(X_A|B) = H(X_A) + \int dx_A \hs p(x_A) S(\rho_B^{x_A}) - S(B)}
we have,
\eqn{K^\rhd &\geq& H(X_A) +  \int dx_A\hs  p(x_A)\hs S(\rho_E^{x_A}) - S(E) - H(X_A|X_B)\nn\\
\ee S(X_A|E) - H(X_A|X_B) \label{DWbound}}
which is what one would expect from the Devetak-Winter relations \cite{Devetak:2005p5086}.
Using the entropic uncertainty relation we can bound the eavesdropper's information on the relevant observable. Substituting from (\ref{uc2}) and recalling that $S(P_A|B)\leq S(P_A|P_B) = H(P_B|P_A)$, we can write,
\eqn{K^\rhd &\geq &\nn \log 4\pi - H(X_A|X_B) - S(P_A|B)\\
\ee \log 4\pi - H(X_A|X_B) - H(P_A|P_B).\label{khappend}}

As discussed in the main text, Alice and Bob can bound their secret key rate for this protocol by measuring Bob's conditional variance. Substituting the Shannon entropy for a Gaussian distribution $H_G(X_B|X_A) = \log\sqrt{2\pi eV_{X_B|X_A}}$, where $V_{X_B|X_A} = V_{X_B} - \EV{X_A X_B}^2/V_{X_A}$, we arrive at a final expression for the key rate of
\eqn{K^\rhd &\geq & \log 4\pi - \log 2 e \pi \sqrt{V_{X_A|X_B}V_{P_A|P_B}} \nn\\
\ee \log \bk{\frac{2}{e\sqrt{V_{X_A|X_B}V_{P_A|P_B}}}}\label{kh}}
The RR expression is obtained by simply permuting the labels of Alice and Bob. That this result is pessimistic is implied by consideration of the predictions for a perfect channel. Under the security proofs presented in \cite{Devetak:2005p5086,GarciaPatron:2006p381,Navascues:2006p805} if Alice and Bob share a pure EPR state (real or effective) with variance $V = \cosh(2s)$ then the key rate given in (\ref{DWbound}) is always positive provided the squeezing parameter is non-zero ($s>0$). On the contrary, the result in (\ref{kh}) is only positive for $V_{X_A|X_B}V_{P_A|P_B} = \mathcal{E}_{\blacktriangleleft} \leq \bk{\frac{2}{e}}^2 \approx 0.55$ which implies a squeezing parameter of $s \geq .15$ or about -1.3 dB of squeezing. This result was also calculated in \cite{agnes}, however as mentioned before the proof relied on the assumption of the applicability of the entropic uncertainty relation to infinite dimensional Hilbert spaces. Furthermore, Ferenczi argued that since for coherent states  the directly measured conditional variances are always greater than 1, the above procedure would never predict a positive key rate for coherent state protocols \cite{agnes}. However, this conclusion comes from a mistaken application of the key rate formulae as we now demonstrate. 

\subsection{Heterodyne-Homodyne (Coherent States and Homodyne Detection)}
\label{sec:co_hom}
Consider a DR coherent state protocol, which in the EB picture involves Alice making a heterodyne detection upon her arm of an EPR pair. Thus she first mixes her mode with vacuum resulting in two modes $A_1$ and $A_2$ upon which she measured $\x$ and $\p$ respectively. To obtain effective EB data from a P\&M scheme Alice will rescale her modulation signal by dividing $\alpha_{x(p)}=+(-)\sqrt{2(V_{x(p)}-1)/(V_{x(p)}+1)}$ where $V_{x(p)} = V_{S_{x(p)}}+1$ and $V_{S_{x(p)}}$ is the modulation variance of the signal encoded in the respective quadratures. Bob then makes a homodyne detection, randomly switching between the quadratures. The DR key rate is then bounded by,
\eqn{K^\rhd & \geq & S(X_{A_1}|E) - H(X_{A_1}|X_B)}
After Alice's projective measurement upon $A_1$ the state $\rho_{A_1BE}$ is pure and we can again apply the entropic uncertainty relation to write,
\eqn{K^\rhd &\geq &\log 4 \pi - S(P_{A_1}|B) - H(X_{A_1}|x_B)\nn\\
&\geq & \log 4\pi - H(P_{A_1}|P_B) - H(X_{A_1}|X_B)}

Now this formula might pose a problem, in that we do not measure $\p$ upon mode $A_1$. Nevertheless, this can be circumvented if we trust the devices, specifically the beamsplitter in Alice's station. As noted earlier, we may safely bound the Shannon entropy by assuming that the statistics are Gaussian and therefore only dependent upon the conditional variance i.e. $H(P_{A_1}|P_B) =  \log\sqrt{2\pi eV_{P_{A_1}|P_B}}$. Then, if we trust Alice's beamsplitter we know that $V_{P_{A_1}|P_B} = V_{P_{A_2}|P_B}$ and thus $H(P_{A_1}|P_B) = H(P_{A_2}|P_B)$ which is directly measured. We therefore have,
\eqn{K^\rhd &\geq & \log 4\pi - \log\sqrt{2\pi e V_{P_{A_2}|P_B}} - \log\sqrt{2\pi e V_{X_{A_1}|X_B}} \nn \\
\ee \log\frac{2}{e\sqrt{V_{X_{A_1}|X_B}V_{P_{A_2}|P_B}}}}

Note that for positive key we now require the condition $V_{X_{A_1}|X_B}V_{P_{A_2}|P_B} \leq 0.55$. To illustrate how correlated a state we now require respect to the purely homodyne protocols we rewrite this as a condition on the homodyne variances.  Assuming quadrature symmetry ($V_{X_A|X_B} = V_{P_A|P_B}$) and using the relationship $2 V_{X_{A_1}|X_B} = V_{X_A|X_B} + 1$ (similarly for $\hat{p}$ quadrature), we find the requirement on the heterodyne-homodyne conditional variance is equivalent to the homodyne-homodyne condition $V_{X_A|X_B}V_{P_A|P_B} \leq 0.22$.

For RR the key rate is bounded by,
\eqn{K^\lhd &\geq & S(X_B|E) - H(X_B|X_{A_1}) \nn \\
&\geq & \log 4\pi - S(P_B|A) - H(X_B|X_{A_1})\nn\\
&\geq &  \log 4\pi - H(P_B|P_A) - H(X_B|X_{A_1})\nn\\
&\geq & \log \frac{2}{e\sqrt{V_{P_B|P_A}V_{X_B|X_{A_1}} }}}

Once again, the quantity $V_{P_B|P_A}$ is not measured directly, but provided we trust Alice's beamsplitter we have $V_{P_B|P_A} = V_{P_B} - 2\EV{P_{A_2}P_B}^2/(2V_{P_{A_2}}-1)$ allowing us to evaluate the bound. However, since we must trust Bob's devices in order to apply the entropic uncertainty relation we explicitly may not trust Alice's devices. Therefore, this RR protocol could never be 1sDI.

\subsection{Homodyne-Heterodyne (Squeezed States and Heterodyne Detection)}
These protocols are essentially mirror images of the coherent state homodyne schemes since, in the EB picture, we now have Alice making homodyne measurements and Bob making heterodyne measurements. We now have Alice swapping between $\x$ and $\p$ measurements while Bob splits up his mode measuring $\x$ upon $B_1$ and $\p$ upon $B_2$. As per the previous squeezed state case, to obtain effective EB data from a P\&M scheme Alice will rescale her modulation signal by dividing $\alpha_{x(p)}=+(-)\sqrt{1-1/V_{x(p)}^2}$. The DR key rate is bounded by,

\eqn{K^\rhd &\geq & S(x_A|E) - H(X_A|X_{B_1}) \nn\\
&\geq & \log 4\pi - S(P_A|B) - H(X_A|X_{B_1}) \nn\\
&\geq & \log 4\pi - H(P_A|P_B) - H(X_A|X_{B_1})\nn\\
 &\geq & \log \frac{2}{e \sqrt{V_{P_A|P_B}V_{X_A|X_{B_1}}}}} 
where we will have to trust the beamsplitter in Bob's station to obtain $V_{P_A|P_B} = V_{P_A} - 2\EV{P_{B_2}P_A}^2/(2V_{P_{B_2}}-1)$  from the directly measured conditional variance.


The RR key rate is bounded by,

\eqn{K^\lhd &\geq & S(X_{B_1}|E) - H(X_{B_1}|X_A) \nn \\
&\geq & \log 4\pi - S(P_{B_1}|A) - H(X_{B_1}|X_{A})\nn\\
&\geq &  \log 4\pi - H(P_{B_1}|P_A) - H(X_{B_1}|X_{A})\nn\\
&\geq & \log \frac{2}{e\sqrt{V_{P_{B_2}|P_A}V_{X_{B_1}|X_{A}} }}}
where we have again used the known action of Bob's beamsplitter to write $V_{P_{B_1}|P_A} = V_{P_{B_2}|P_A}$. Similarly to previous section, only the RR protocol is 1sDI.

\subsection{Heterodyne-Heterodyne (Coherent States and Heterodyne Detection)}
The final protocols involve Bob making a heterodyne measurement upon coherent states, or alternatively both parties making heterodyne measurements upon a two-mode squeezed vacuum. We include this for completeness, however none of these protocols could be 1sDI because devices on both sides must always be trusted. Thus there are now four modes involved $A_1$ and $B_1$ upon which $\x$ is measured and $A_2$ and $B_2$ upon which $\p$ is measured. We can consider the $\x$ and $\p$ channels separately. Note that this is actually an underestimation of the key rate as it essentially allowing Eve to devote all her resources to estimating either the $\x$ or $\p$ measurements separately, whereas in reality she must in fact estimate both simultaneously. As per the previous coherent state case, to obtain effective EB data from a P\&M scheme Alice will rescale her modulation signal by dividing $\alpha_{x(p)}=+(-)\sqrt{2(V_{x(p)}-1)/(V_{x(p)}+1)}$. The DR key rate for $\x$ is bounded by,
\eqn{K^\rhd &\geq & S(X_{A_1}|E) - H(X_{A_1}|X_{B_1}) \nn\\
&\geq & \log 4\pi - S(P_{A_1}|B) - H(X_{A_1}|X_{B_1})\nn\\
&\geq & \log 4\pi - H(P_{A_1}|P_B) - H(X_{A_1}|X_{B_1})\nn\\
&\geq & \log \frac{2}{e\sqrt{V_{P_{A_1}|P_B} V_{X_{A_1}|X_{B_1}}}}}
Provided we trust the beamsplitter in Bob's station we can have $V_{P_{A_1}|P_B} = V_{P_{A_2}|P_B}= V_{P_{A_2}} - 2\EV{P_{A_2}P_{B_2}}^2/(2V_{P_{B_2}}-1)$.

The RR key rate is bounded by,
\eqn{K^\lhd &\geq & S(X_{B_1}|E) - H(X_{B_1}|X_{A_1}) \nn\\
&\geq & \log 4\pi - S(P_{B_1}|A) - H(X_{B_1}|X_{A_1})\nn\\
&\geq & \log 4\pi - H(P_{B_1}|P_A) - H(X_{B_1}|X_{A_1})\nn\\
&\geq & \log \frac{2}{e\sqrt{V_{P_{B_1}|P_A} V_{X_{B_1}|X_{A_1}}}}}
By trusting the beamsplitter in Alice and Bob's station we get $V_{P_{B_1}|P_A} = V_{P_{B_2}|P_A}= V_{P_{B_2}} - 2\EV{P_{B_2}P_{A_2}}^2/(2V_{P_{A_2}}-1)$.

\section{Security proof with imperfect reconciliation efficiency}

In the main text we derived secret key rates assuming that Alice and Bob achieve the Shannon capacity for their Gaussian encoding. Thus the key rate is bounded by,
\eqn{K^\rhd \geq I(X_A:X_B) - \chi(X_A:E)}
Here we now use $X_A$ to denote that fact that depending upon the particular protocol we could be referring to quadrature measurements with or without a shot noise penalty and also that these quantities must be averaged over the $\x$ and $\p$ quadratures in  the case that the two are not perfectly symmetric.
In reality we won't be able to perfectly achieve this capacity and the key rate will instead be bounded by,
\eqn{K^\rhd \geq \beta I(X_A:X_B) - \chi(X_A:E)}
where $\beta <1$ is the reconciliation efficiency. In this case, instead of using the entropic uncertainty relation to lower bound the secret key rate, we will use it to upper bound Eve's information and then independently measure $\beta I(X_A:X_B)$ to obtain the actual key rate.
Eve's information is upper bounded by the Holevo quantity thus,
\eqn{\chi(X_A:E) \leq S(E) - \int dX_A \hs p(X_A) S(\rho_E^{X_A})}
The conditional von Neuman entropy of the observable $X_A$ is given by 
\eqn{S(X_A|E) = H(X_A) + \int dX_A \hs p(X_A) S(\rho_E^{X_A}) - S(E)}
Thus we can rewrite Eve's information as,
\eqn{\chi(X_A:E) \leq H(X_A) - S(X_A|E)}
We now make use of our CV entropic uncertainty relation,
\eqn{S(X_A|E) + S(P_A|B) \geq \log 4\pi}
to obtain,
\eqn{\chi(X_A:E) \leq H(X_A) + S(P_A|B) -\log 4\pi}
Using the fact that $S(P_A|B) \leq S(P_A|P_B) = H(P_A|P_B)$ and that the Shannon entropy is maximised by a Gaussian distribution for a fixed variance such that $H(X_A) \leq \log \sqrt{2\pi e V_{X_A}}$ we arrive finally at,

\eqn{\chi(X_A:E) \leq \log 2\pi e \sqrt{V_{X_A}V_{P_A|P_B}}-\log 4\pi}
Thus the secret key rate for an arbitrary $\beta$ is ,
\eqn{K^\rhd \geq \beta \log \sqrt{\frac{V_{X_A}}{V_{X_A|X_B}}} 
+\log \frac{2}{ e \sqrt{V_{X_A}V_{P_A|P_B}}} \label{beta9}}
The RR key rate is given by interchanging Alice and Bob to obtain,
\eqn{K^\lhd \geq \beta \log \sqrt{\frac{V_{X_B}}{V_{X_B|X_A}}} 
+\log \frac{2}{ e \sqrt{V_{X_B}V_{P_B|P_A}}} \label{beta10}}
Finally, the key rate for the RR protocol where Bob heterodynes is given by
\eqn{K^\lhd \geq \beta \log \sqrt{\frac{V_{X_{B_1}}}{V_{X_{B_1}|X_A}}} 
+\log \frac{2}{ e \sqrt{V_{X_{B_1}}V_{P_{B_2}|P_A}}} \label{beta11}}
and the DR protocol where Alice heterodynes (or alternatively prepares coherent states) is given by
\eqn{K^\rhd \geq \beta \log \sqrt{\frac{V_{X_{A_1}}}{V_{X_{A_1}|X_B}}} 
+\log \frac{2}{ e \sqrt{V_{X_{A_1}}V_{P_{A_2}|P_B}}} \label{beta12}}

\section{Experimental Details and modeling of EB scheme with Homodyne-Homodyne detection }
\label{sec:exp_eb_hh}
\begin{figure*}[t]
\centering
\includegraphics[width=\linewidth]{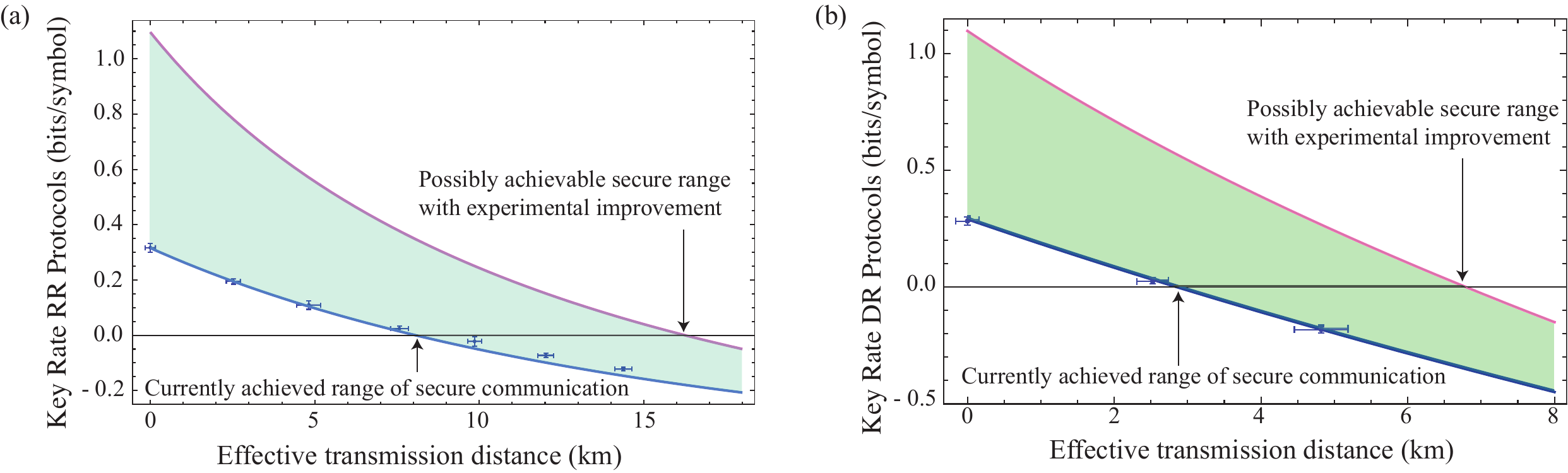}
\caption{Predicted improvement of secure transmission distance through optical fibre for the EB protocols with an improved experimental setup (red curve) for the (a) RR protocol and (b) DR protocol. The model for the current system (blue curve) is plotted along with experimental data (blue points) for comparison. The two OPAs in our system each produced -6 dB of squeezing and 10.7 dB of anti-squeezing and suffered from 13\% combined loss due to the squeezers' cavities and the propagation through the optical components. A further loss of 6\% and 8\% was due to the inefficiency of Alice and Bob's homodyne detections respectively. The value of the unknown rotation, $\theta$, was estimated to be $3\pi/180$. The improved system consists of two squeezers each producing -10 dB of measurable squeezing and 16 dB of anti-squeezing, 5\% loss due to the cavities and propagation of the optical beams through the optical components with 4\% and 5\% loss for Alice and Bob's stations respectively and a rotation of $\pi/120$. Reconciliation efficiency is chosen to be 0.95 for both cases. These theoretical lines are produced using the model described in the section~\ref{sec:exp_eb_hh} and equations (\ref{beta9}) and (\ref{beta10}).}
\label{fig:Homideal}
\end{figure*}

In this section, we discuss the experimental details, the imperfections and the modelling of the experiment with EB source and homodyne-homodyne measurements. 
A 1064 nm Nd:YAG laser source which was frequency doubled to 532 nm was used in the experiment. Both light fields were passed through mode cleaning cavities for spatial filtering and also to provide quantum noise limited coherent beams. Two similar degenerate bow-tie optical parametric amplifiers were used to produce two amplitude squeezed states. The 1064 nm field was used as the seed field for the OPAs and also as the local oscillator for homodyne detections. The 532 nm field was used to pump the OPAs. Here, we used PPKTP crystal to produce nonlinear effects. 
The estimated values of squeezing and anti-squeezing generated from both OPAs were -6 dB and 10.7 dB at 3 MHz. A simple model was used to infer the pure squeezing produced inside the cavity and also the effective loss of the system. According to this model, the OPAs were assumed to produce pure squeezed states and the effective loss, the combination OPA's propagation and detectors' losses, was modelled with a beam splitter after the squeezer. Using this model and the measured values of the squeezing and anti-squeezing, we predicted that each OPA produced 11.5 dB of pure squeezing (-11.5 dB of squeezing and 11.5 dB of anti-squeezing), and the effective loss of this system was calculated to be around 20\% with 14\% loss due to each squeezer's cavity and also the propagation of the optical beams through the optical components and nearly 6\% loss due to each detection station.  
 
Four identical photodiodes were used in the detection stations. The detection efficiency of Alice and Bob’s stations were estimated to be 94\% and 92\% respectively, with fringe visibility of 99\% and the photodiodes’ quantum efficiency of around 96\% for all the detectors. We estimated 2\% extra loss on Bob’s side due to the loss introduced by the half wave plate and polarizing beamsplitter that were used to simulate the lossy channel. Each pair of detectors were balanced electronically,
providing 30 dB of common mode rejection. Each detector had at least 16 dB of dark noise clearance.

For each separate homodyne detection $5\times10^7$ data points are sampled at $14\times10^6$ samples per second. In order to provide sufficient statistics for each data point, this process is repeated ten times. These data were then digitally filtered to 2.5-3.5 MHz and then resampled. After this process, the number of data points was reduced to $4\times10^6$ which is sufficient to extract the key rates. 

Other experimental imperfections, such as the  the imperfect locking points and the unbalanced beamsplitter ratios, were also captured in the model. Since all the states and operators were assumed to be Gaussian in this experiment, the states can be easily described by their mean values and covariance matrices (CM's). The effect of Gaussian operations on Gaussian states can be compactly calculated via symplectic transformations \cite{Weedbrook:2012p5160}. Under an arbitrary symplectic operation, $S$, an input CM, $\gamma_{\mathrm{in}}$ transforms via 
\eqn{\gamma_{\mathrm{out}} = S \gamma_{\mathrm{in}} S^T \label{cmtrans}}
\\The CM of a two mode squeezed vacuum with squeezing in quadrature in modes $i$ and $j$ is given by applying the following symplectic operator,

\quad
$$SQ_{i,j}(s_1,s_2)=
\begin{pmatrix} 
e^{s_1} & 0 & 0 & 0 \\
0 & e^{-s_1} & 0 & 0\\
0 & 0 & e^{-s_2} & 0\\
0 & 0 & 0 & e^{s_2}
\end{pmatrix}
\quad
$$
where $s_1$ and $s_2$ are squeezing parameters applied on the $i^{\mathrm{th}}$ and $j^{\mathrm{th}}$ second mode respectively. Implicit in this notation is the fact that when applied to a multi-mode CM one should appropriately pad out the above matrix such that the identity is applied to all modes other than $i$ and $j$.

The loss of each squeezer is modelled by introducing a vacuum mode, and then applying a beamsplitter of transmittance $\eta_{A(B)}$ on each squeezed mode and a vacuum mode to mix them together. The beamsplitter transformation between the modes $i$ and $j$ is: 

$$BS_{i,j}(\eta)=
\begin{pmatrix} 
\sqrt{\eta} & 0 & -\sqrt{1-\eta} & 0  \\
0 & \sqrt{\eta} & 0 & -\sqrt{1-\eta}  \\
\sqrt{1-\eta} & 0 & \sqrt{\eta} & 0 \\
0 & \sqrt{1-\eta} & 0 & \sqrt{\eta}  \\

\end{pmatrix}
\quad
$$

In order to create an EPR state two squeezed states are locked in quadrature and mixed on a 50:50 beamsplitter. To model the imperfect locking point a phase shift $\theta$ is applied to one mode before they mix on a beamsplitter. The applied operator is as follows:
$$RT_i(\theta)=
\begin{pmatrix} 
\cos\theta & -\sin\theta  \\
\sin\theta & \cos\theta \\
\end{pmatrix}
\quad
$$
To model the loss of the transmission channel, a vacuum state was introduced and mixed with the second mode on a beamsplitter with transmittance $T$. The loss of each homodyne station was modelled by a beamsplitter of transmittance $\eta_{D_{A(B)}}$, equal to the efficiency of the homodyne station, with the other mode being in a thermal state of variance $V_{\Delta_{A(B)}} = 1+ \Delta_{A(B)}/(1-\eta_{D_{A(B)}})$ to model the detector dark noise of magnitude $\Delta_{A(B)}$. 
Thus the final CM is given by (\ref{cmtrans}) with,
\eqn{S \ee BS_{2,7}(\eta_{D_B})BS_{2,6}(T)BS_{1,5}(\eta_{D_A})BS_{1,2}(1/2)RT_2(\theta)\nn\\
&&BS_{2,4}(\eta_B)BS_{1,3}(\eta_A)SQ_{1,2}(s_1,s_2) \label{Strans}}
with
\eqn{\gamma_{\mathrm{in}} = \mathrm{diag}(1,1,1,1,1,1,1,1,V_{\Delta_A},V_{\Delta_A},1,1,V_{\Delta_B},V_{\Delta_B)}}
a 14x14 diagonal matrix.
To determine the value of the applied loss, $T$, from the measured correlations it is sufficient to consider the ratio of the correlation between Alice and Bob at particular loss setting with the case where the channel is set to full transmission. Using equations (\ref{beta9}) and~(\ref{beta10}), key rates were calculated from both the simulated covariance matrix and experimentally measured conditional variances. We plotted them as a function of the effective transmission distance, showing excellent agreement with the experimental results as shown in Fig \ref{fig:Homideal}. 

The model was also used to estimate the performance of a more efficient system with two squeezers each producing -10 dB of squeezing and 16 dB of anti-squeezing and detection efficiency for Alice and Bob’s stations of 96\% and 95\% respectively. Using these parameters, our model shows that the range of the secure communication would extend from 7.5 km to 15 km for the RR protocol and from 2.5 km to more than 6 km for DR protocol. If the phase shift, $\theta$, which was used to model the imperfect locking point of the EPR state reduced from $3\pi/180$ to $\pi/120$, our model predicts that the range of the secure communication would extend further to 17 km for the RR protocol (Fig. \ref{fig:Homideal}(a)) and to more than 8 km for DR protocol (Fig. \ref{fig:Homideal}(b)). Achieving this level of quadrature squeezing and phase stability are experimentally challenging but feasible as up to -12 dB of squeezing was reported previously \cite{Roman.Schnabel:2010p251102}.

\begin{figure*}[t]
\centering
 \includegraphics[width=0.6\linewidth]{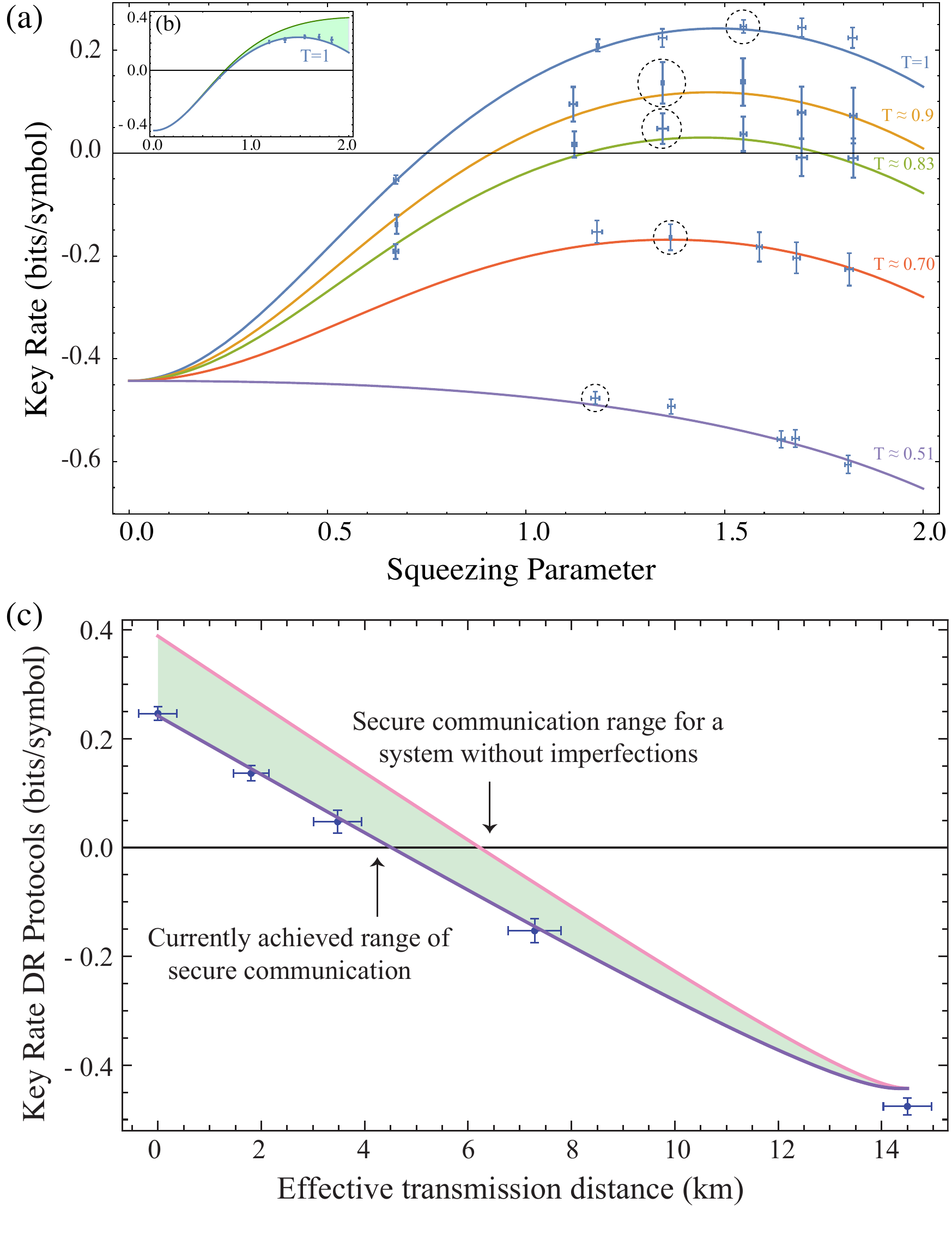}
\caption{(a) Variation of key rates versus effective modulation squeezing parameter for 5 different transmissions. A theory line with the average transmission of the channel is fitted on the experimental data points with 1 s.d. error bars. Data points surrounded by dashed circles correspond to the optimum modulation squeezing parameters which result in the highest key rate for each transmission. The key rates resulting from these optimum modulation variances are shown separately in (c). Inset (b) demonstrates the gap between the theoretical cross-talk free model and the realistic model which captures the experimental imperfections. (c) Predicted improvement of secure transmission distance through the optical fibre for the coherent state protocol with an improved experimental setup (red curve). The model for the current system (blue curve) is plotted along with experimental data (blue points) for comparison. In the actual experiment, the optimal modulation variance is reduced due to unwanted cross-quadrature correlations. In the improved setup, the cross-talk has been eliminated and the optimal modulation variance is now determined by the reconciliation efficiency, which is chosen to be 0.95 for both cases. These theoretical lines are produced using the model described in section~\ref{sec:coh_pm}, where the value of the unknown rotation, $(\theta_x,\theta_p)$, was estimated to be about $\approx (6\pi/180,3\pi/180)$ and equation (\ref{beta12}). 
}
\label{fig:SQ}
\end{figure*}  

\section{Experimental Details and modeling of P\&M with coherent states and homodyne detection}
\label{sec:coh_pm}
In this section, we discuss the experimental details, imperfections and modelling of the experiment with coherent states and homodyne measurements. 
A quantum noise limited 1064 nm laser was used in the experiment. A small portion of it was passed through a pair of phase and amplitude electro-optic modulators (EOMs). EOMs were used to provide a Gaussian distributed modulation on each quadrature. Each EOM was driven by an independent function generator, providing a broadband white noise signal up to 10 MHz. The magnitude of white noise was set to provide almost the same displacement on each quadrature. Outputs of function generators were divided into two. One part was sent to drive the EOMs and the other was recorded. This modulation record, after calibration, was Alice's data since she had control over the source. Here, calibration means determining the relationship between the function generator output and the phase space displacement as measured before transmission. We denote the variance of this phase space modulation $V_S$. The modulated beam was then sent through a lossy channel to Bob. To model the lossy channel, a vacuum state was introduced to the system and was mixed with the Bob's mode on a beamsplitter of transmission $T$. Upon receiving his mode, Bob performed a homodyne measurement, alternating between conjugate quadratures. An electronic delay was introduced to Alice's and Bob's data to gain the maximum correlation between them at 3.5-4.5 MHz.

When the homodyne detector was locked to the phase quadrature there was 30 dB suppression of cross correlation between orthogonal quadratures and when it was locked to the amplitude quadrature the suppression was around 37 dB. Our pair of detectors, both with dark noise clearance of 18 dB, were balanced electronically, providing 30 dB of common mode rejection. Our homodyne efficiency was around 95\% with fringe visibility of 98\%, limited by the mode distortions introduced by the EOMs. The photodiode's quantum efficiency was estimated to be around 98.5\%.  $4\times10^6$ data points were sampled at $25\times10^6$ samples per second utilizing a digital data acquisition system. The process was repeated five times in order to provide sufficient statistics for each data points. These data were then digitally filtered to 3.5-4.5 MHz.

In order to find the maximum range over which the protocol provides secure communication, we wish to find the optimal modulation variance for each value of the channel transmission. We scanned the modulation variance over a range of 2 to 19 times the shot noise. As discussed in section~\ref{sec:co_hom}, by rescaling Alice's recorded signal, the key rates can be calculated using 
~\eqref{beta12} with reconciliation coefficient set to 0.95.

To highlight the relative advantages of the coherent state source, consider the CM of the equivalent two mode EPR state:
\eqn{\mathrm{EPR}(s)=
\begin{pmatrix} 
\cosh(2s) & 0 & \sinh(2s) & 0 \\
0 & \cosh(2s) & 0 & -\sinh(2s)\\
\sinh(2s) & 0 & \cosh(2s) & 0\\
0 & -\sinh(2s) & 0 & \cosh(2s)
\end{pmatrix}
\quad
\label{eprcm}}
where \textit{s} is the squeezing parameter which is related to the modulation variance via $\cosh(2s) = V_S+1$. To model the prepare \& measure experiment, we remain in the equivalent EB picture and begin with $\gamma_{\mathrm{in}} = \mathrm{EPR(s)}$. Recall that one part of the equivalent EPR state was sent to Bob through a lossy channel where he performed a homodyne detection, and on the other part Alice performed a heterodyne detection. Although much more flexible, the coherent state setup naturally still suffers from imperfections which in turn effect the optimum modulation. These imperfect correlations arise partly from cross correlation between orthogonal quadratures and partly from our limited ability to maximize the correlation between Alice and Bob's modes using electronic delay. Both phenomena can be thought of as an unknown rotation in the system. A rotation operator with small angles is applied to the \textit{X} and \textit{P} quadratures of the second mode (Bob's mode) to model the imperfect correlation between Alice's and Bob's modes.

The channel transmission, $T$, can again be determined directly by taking the ratio of the correlation at a particular setting with the correlation at full transmission. Technically, the experimental channel would also introduce a small amount of excess noise, however this is negligible compared to the excess noise coming from the effects described above. The final simulated covariance matrix hence is \eqn{\gamma_{\mathrm{out}} = S [\gamma_{\mathrm{in}}\oplus V_\chi(B) \oplus \mathrm{diag(1,1)}] S^T,}
where $S$ is given by
\eqn{S =RT(\theta_x,\theta_p)BS_{1,4}(1/2) BS_{2,3}(T).}
Here, $V_\chi=\textrm{diag}(1+\chi_x,1+\chi_p)$, where $\chi_{x(p)}$ is the excess noise in $\hat{x}(\hat{p})$ quadrature. The rotation matrix, \eqn{RT(\theta_x,\theta_p)=\begin{pmatrix} 
\cos\theta_x & \sin\theta_x  \\
-\sin\theta_p & \cos\theta_p \\
\end{pmatrix}} serves as the fitting parameter to model the unknown rotation due to aforementioned experimental imperfection.

To evaluate the key rate~\eqref{beta12}, we neglect the excess noise in the channel by setting $\chi_{x(p)}=0$. The variation of key rates versus the equivalent modulation squeezing parameter for 5 different transmissions is shown in Fig.~\ref{fig:SQ}(a). As the modulation is increased, so too is the detrimental effect on the correlations, leading to a smaller value for the optimal modulation parameter, whereas for an ideal experiment, this would depend only upon $\beta$. In inset(b) of Fig.~\ref{fig:SQ} the gap between the ideal case without cross correlation and the realistic case is shown for the case of perfect transmission. For each transmission value, the modulation squeezing parameters that provide the highest key rate are chosen and plotted in Fig.~\ref{fig:SQ}(c). These optimum variances match well with what our theoretical model predicts. As is clear from Fig.~\ref{fig:SQ}(a), using coherent states provides a much greater range over which to tune the equivalent squeezing. When using actual EPR states the maximum achievable value for $s$ is around 0.8, well short of the optimum. Our model also predicts that if the cross correlation between Alice and Bob's modes was zero, the range of secure communication for this protocol would extend from 4.5 km to 6.5 km as depicted in Fig. \ref{fig:SQ}(c).  

\end{appendix}




\end{document}